\documentclass[conference]{IEEEtran}
\IEEEoverridecommandlockouts

\usepackage{cite}
\usepackage{amsmath,amssymb,amsfonts}
\usepackage{algorithmic}
\usepackage{graphicx}
\usepackage{textcomp}
\usepackage{xcolor}
\def\BibTeX{{\rm B\kern-.05em{\sc i\kern-.025em b}\kern-.08em
		T\kern-.1667em\lower.7ex\hbox{E}\kern-.125emX}}
\newcommand{\revision}[1]{\textcolor{black}{#1}}
\begin{document}
	
	\title{Accelerating Hardware Verification with Graph Models
		
	}

   \author{%
   	\IEEEauthorblockN{%
   		Raghul Saravanan, Sreenitha Kasarapu, Sai Manoj Pudukotai Dinakarrao
   	}%
   	\IEEEauthorblockA{ Department of Electrical and Computer Engineering, George Mason University, Fairfax, VA, USA}%
   	{rsaravan@gmu.edu, skasarap@gmu.edu, spudukot@gmu.edu}
   	
    }

	\maketitle
	
	\begin{abstract} 
		
	The ever-expanding intricacies within the architectural design spectrum of modern processors and intellectual property (IP) designs foster a challenging environment for identifying and mitigating potential flaws in the early stage of the IC design cycle. Ensuring system integrity and security amidst this complexity demands continuous innovation in design verification methodologies.

	
	In the era of hardware design verification, Hardware Fuzzing -- a robust technique inspired by software testing has recently gained traction for its effective bug detection in complex hardware designs. Lately, hardware fuzzing frameworks are emerging and successfully identifying bugs in cutting-edge processors. Detecting the bugs and vulnerabilities demands a complex network of instruction seeds to ensure thorough coverage of intricate design elements of the CPU. However, hardware bugs can manifest at different abstraction levels, including RTL (pre-synthesis), gate-level netlist (post-synthesis), and physical layout. Comprehensive verification across all levels is crucial to ensure security and functionality. Gate-level verification is particularly important, as it can uncover bugs introduced during synthesis that are missed at higher abstraction levels. However, the complexity of gate-level netlists in modern SoCs poses significant challenges, leading to longer simulation times and resource-intensive verification.
	
	In this paper, we propose a novel hardware fuzzer, \textit{GraphFuzz}, a graph-based hardware fuzzer to overcome the limitations and enhance the fuzzing process at the gate-level netlist. In this approach, hardware designs are modeled as graph nodes, where intrinsic hardware behaviors of the gates are encoded as features. By applying graph learning algorithms, we analyze node patterns and detect potential vulnerabilities effectively. We test our fuzzer on varied hardware designs, including benchmark circuits, peripherals, and intricate design elements of popular open-source processors. Our graph model representing the netlist during fuzzing achieves an average accuracy of 80\% and detects bugs with an average accuracy of 70\%  demonstrating the effectiveness of using graph-based techniques to enhance hardware verification and identify potential vulnerabilities at the gate-level netlist.

	\end{abstract}
		\section{Introduction}
	\label{intro}

	The ever-increasing threat landscape of hardware designs in the modern IC design flow, exacerbated by global collaboration between vendors, necessitates sophisticated hardware verification mechanisms to ensure system integrity \cite{icdesignflow1}. As the complexity of integrated circuits (ICs) and the reliance on heterogeneous third-party intellectual property (IP) blocks—especially in system-on-chips (SoCs)—grows, so do the vulnerabilities that can be exploited \cite{apple}. These security vulnerabilities arise not only from the involvement of third-party vendors but also throughout the design, implementation, and integration phases of the modern IC design flow, posing significant challenges to SoC security \cite{Artenstein'17, Liu'17, Lipp'18, Kocher'18}. Such vulnerabilities can manifest in various forms, including information leakage, access control violations, side-channel attacks, and the insertion of hardware Trojans \cite{Bulck'18,Schwarz'19,Jang'16,Wojtczuk'12,Feng'22,Ghosh'23,Kang'19,Alatoun'21,Tang'22,Mohandoss'18,Vangal'08}. Furthermore, existing countermeasures such as logic locking, camouflaging, and watermarking are proving inadequate in addressing these threats due to the increasing complexity of large designs \cite{watermarking,Sasan21,mtswarup,Rajendran2013SecurityAO}. 
	Unlike software, hardware patches are costly and not practical, making it crucial to address vulnerabilities before fabrication. For instance, Intel’s machine check bug, for instance, enabled denial-of-service attacks, costing nearly half a billion dollars in re-engineering.

	Security vulnerabilities stem from the underlying hardware, such as RTL and gate-level netlists, necessitating a concentrated focus on hardware verification at the logic level.  To nullify the effects of the vulnerabilities and bugs in the SoCs and ICs, hardware verification community has developed various pre-silicon verification techniques \cite{Chen'11, Mukherjee'15, Dessoky'19, Sarangi'06, Wagner'07,  Jin'05, Gogri'22, Guzey'10} and Electronic Design Automation (EDA) tools \cite{Siemens-Questa, SymbiYosys, ABC}. Formal methods and information flow tracking (IFT) are the most popular verification techniques. However, recent advancements in formal methods, such as satisfiability checking (SAT) \cite{Azar2018SMTAN, smt} and satisfiability modulo theories (SMT) \cite{model}, have encountered significant challenges due to scalability limitations and the requirement for expert knowledge. These constraints diminish their effectiveness in verifying increasingly complex modern hardware designs. Furthermore, while Information Flow Tracking (IFT) techniques \cite{ift1, ift2, ift3} are effective in tracking sensitive data flows, they require design instrumentation, which in turn demands additional computational time and resources, further complicating their application to large-scale systems. There is a growing need for hardware verification methodologies and frameworks for modern SoCs that can scale to large, complex designs \cite{Wang'18, Tiwari'11, Ardeshiricham'17, Li'11, Li'14, Zhang'15, Meng'22}.

	To address the shortcomings of the existing pre-silicon verification, recently hardware community researchers introduced hardware fuzzing inspired from software testing paradigm \cite{Laeufer'18,Li'21, Hur'21,Trippel'22,Muduli'20, Kabylkas'21, Kande'22, rsglsvlsi, vnp, rsglsvlsi2, saravanan2024emergencehardwarefuzzingcritical}. Fuzzing is a widely used software testing technique for bug detection in software applications. 
	Industry-based fuzzing frameworks such as Google's OSS-Fuzz \cite{Serebryany'17} and Microsoft's Security Risk Detection \cite{microsoftrisk} have proved their efficacy and effectively identified a plethora of security vulnerabilities. The popular software fuzzer, American Fuzzy Lop (AFL) \cite{AFL'23}, is predominantly used in software fuzzing and is predominantly used in majority of the software fuzzing techniques.

	The initial work on hardware fuzzing, RFuzz \cite{Laeufer'18} adopted software fuzzing methodologies directly onto hardware, while the latter works translated the hardware to an equivalent software and the fuzzing the resultant software code \cite{Trippel'22}. Unfortunately, these  fuzzing methods inherently fail to capture the intrinsic hardware behaviors including logic transitions, and register computations. To circumvent the aforementioned challenges, recent works have proposed fuzzing the hardware at its native hardware abstraction level, making it compatible with traditional hardware verification techniques \cite{Kande'22,hybrid, Hur'21, Canakci'23}. These existing works have proven efficient in identifying bugs and vulnerabilities in state-of-the-art open-source CPU RTL designs \cite{ariane,RISC-V,RISC, Morlkx}, highlighting the prominence of fuzzing at hardware level abstraction \cite{sigfuzz,Kande'22, hybrid, Canakci'23}.

	To this end, there have been several works on hardware fuzzing \cite{Laeufer'18, Kande'22, hybrid, Trippel'22, Hur'21, direct, hyperprop} that aim to fuzz hardware at the RTL abstraction level, focusing on verifying the design's adherence to its specification. \textit{However, it is important to note that hardware vulnerabilities can manifest at different levels of hardware abstraction \cite{property}, including the RTL level (pre-synthesis), gate-level netlist (post-synthesis) and the physical level (layout)}.  To ensure comprehensive security and functionality, hardware verification must be performed at every abstraction level of the design flow. As we elaborate in Section \ref{motivation}, verification at the gate-level is critical as it can capture bugs and vulnerabilities introduced during the synthesis that go unnoticed in higher level of abstraction, ensuring that the final implementation adheres to the intended behavior specified at the RTL level before entering physical design phase. 
	
	Conversely, verification at the gate-level netlist, which encompasses thousands or even millions of gate components in complex SoC designs, poses significant challenges \cite{Farzana'19}. The sheer complexity of the design at this level leads to a substantial increase in simulation time, making the verification process slower and resource-intensive compared to higher abstraction levels.  
	Traditional simulation techniques may struggle to efficiently handle such large-scale designs, especially when running exhaustive test cases. Therefore, there is a compelling need for accelerated hardware verification techniques specifically tailored for the gate-level netlist. Innovative approaches leveraging parallelization, or machine learning (ML) can help address these challenges and ensure that the gate-level verification process remains scalable and effective.

	\paragraph{\textbf{Proposed Work and Key Contributions:}} In our current work GraphFuzz, we propose a novel graph-based hardware fuzzing approach aimed at accelerated hardware verification at the gate level (Section \ref{proposed}), designed to address the challenges and limitations (Section \ref{motivation} ). \textit{To the best of our knowledge, this is the first work on hardware fuzzing at gate-level netlist using graph learning.} In our approach, the gate-level netlist is represented as a graph (Section \ref{graphgen}), where each node and edge encapsulates the inherent hardware features of the design (Section \ref{nodeencode}). We leverage state-of-the-art Graph Learning techniques (Section \ref{grnn}) to train models on this representation using inherent EDA datasets (Section \ref{edadataset}), effectively capturing the behavior of the netlist. This allows for more accurate modeling of the gate-level netlist, enhancing the fuzzing process and improving the overall efficiency of hardware verification. Following the learning of the structural and functional characteristics of the gate-level netlist, the graph, henceforth referred to as NetGraph, undergoes fuzzing via the NetGraph Fuzzer (Section \ref{graphfuzzer}) leveraging the graph inferencing, which applies input directly to the graph at the gate-level netlist abstraction. This approach accelerates the verification process, enabling faster detection of discrepancies between the RTL and gate-level netlist. On the whole, our GraphFuzz offers the following key benefits: 1) supports conventional hardware design and verification methods, 2) captures intrinsic hardware characteristics using comprehensive EDA data, 3) accelerates hardware verification at the gate-level, 4) effectively detects bugs at the gate-level, and 5) does not require extensive design knowledge or expertise in hardware design.  The cardinal contributions of the proposed GraphFuzz are outlined below: 

	\begin{itemize}
		\item We propose a novel hardware fuzzer, GraphFuzz, that leverages state-of-the-art graph learning algorithms to accelerate hardware verification at the netlist level by representing the gate-level netlist as a graph (Section \ref{proposed}, \ref{graphgen}, \ref{nodeencode}). 
		\item GraphFuzz introduces a specialized graph learning algorithm  (Section \ref{grnn}), along with NetGrapher (Section \ref{graphfuzzer}) leveraging graph inference algorithm, which enables efficient fuzzing of the netlist graph for faster bug detection.
		\item We present a method to generate datasets corresponding to EDA tools to improve machine learning model performance (Section \ref{edadataset}). A key component of our work is the creation of a dedicated dataset designed to enhance the performance of machine learning models in the hardware verification context. By curating and optimizing the dataset for machine learning, we ensure that the models used in the NetGraph Fuzzer are trained with high-quality, representative data.
		\item To demonstrate the effectiveness of the NetGraph Fuzzer, we conducted extensive evaluations on several industry-standard benchmarks. These include the ISCAS benchmarks , commonly used to test the functionality of hardware verification tools, as well as IP peripherals such as AES \cite{opentitan} and DSP \cite{opentitan}, and sub-components of real-world open-source CPU designs: 1) or1200 processor (OpenRISC ISA) \cite{RISC}, 2) mor1kx processor (OpenRISC ISA) \cite{RISC}, 3) Ariane processor (a.k.a. CVA6) (RISC-V ISA) \cite{ariane}, 4) IBEX (RISC-V ISA) \cite{RISC-V}, and 5) Rocket (RISC-V ISA) \cite{RISC-V}. 
		
	\end{itemize}
	
	\section{Background}
	\label{background}
	In this section, we provide a brief overview of the modern Integrated Circuit (IC) design and verification process (Section \ref{hwdevpipeline}), followed by an introduction to hardware fuzzing (Section \ref{hwfuzz}). We also emphasize the importance of hardware verification at the post-synthesis level (Section \ref{motivation}) (commonly referred to as gate-level netlist verification) which is our primary motivation for our proposed work (Section \ref{proposed}). Finally, we discuss why graph learning is a promising approach for accelerating hardware fuzzing, especially at the gate-level (Section \ref{gleda}).

	\subsection{The Hardware Development Pipeline} \label{hwdevpipeline}
	
	The design of modern ICs has become increasingly complex due to advancements in technology and the growing demands for performance, power efficiency, and security.  The design process commences with the formulation of system-level requirements and specifications, which outline the functional, performance, and security needs of the IC. HW Design engineers design the IC at the RTL level using hardware description languages (Verilog, VHDL, System Verilog),  where functionality is described in terms of data flow between registers and logic operations during clock cycles. Before entering the next phase, the RTL design undergoes extensive verification to ensure that it meets the functional requirements. The verification is typically done through RTL simulation via the EDA tools ,  where test vectors are applied to check the design's correctness and is free from functional bugs. 
	
	Moving forward, EDA tools synthesize the RTL design and translate the designs to gate-level designs. During this process, the RTL design is translated and mapped onto basic hardware components, such as Boolean gates, flip-flops, buffers, and other standard cells, which are part of the Standard Technology Cell Library provided by the semiconductor foundry. After synthesis, verification at the gate-level netlist is essential to ensure that the hardware implementation matches the functional and timing requirements \cite{Farzana'19}. Verification engineers use formal, functional, simulation, fault simulation-techniques to check if the design meets the RTL specifications \cite{Rajendran2013SecurityAO}. Post-synthesis verification is much more detailed but also much slower compared to RTL verification.
	
	Finally, the design enters the physical design phase, where EDA tools convert the gate-level netlist to a layout, followed by physical verification, where it is prepared for tape-out and sent to the foundry for manufacturing. To this end, all the verification methodologies are pre-silicon, where the model of the design is RTL, gate-level, or layout, rather than an actual silicon artifact. One advantage of pre-silicon verification is the high observability, meaning that any signals within the design can be easily observed and verified. This allows for detailed monitoring and debugging, which is more challenging in post-silicon validation. Conversely, after the chip is manufactured, it undergoes post-silicon verification. Post-silicon verification methods are complex and expensive, offering less observability compared to pre-silicon stages. The primary goal of post-silicon verification is to ensure that the silicon manufacturing process has produced functioning chips that meet the design specifications and perform as intended. During the verification process at each stage, the adjacent levels are compared against their respective Golden Reference Model (GRM) for validation.
	
	\subsection{Hardware Fuzzing} \label{hwfuzz}

\begin{figure}[htb!]
		\vspace{-1em}
		\centering
		\includegraphics[width=1\linewidth]{
			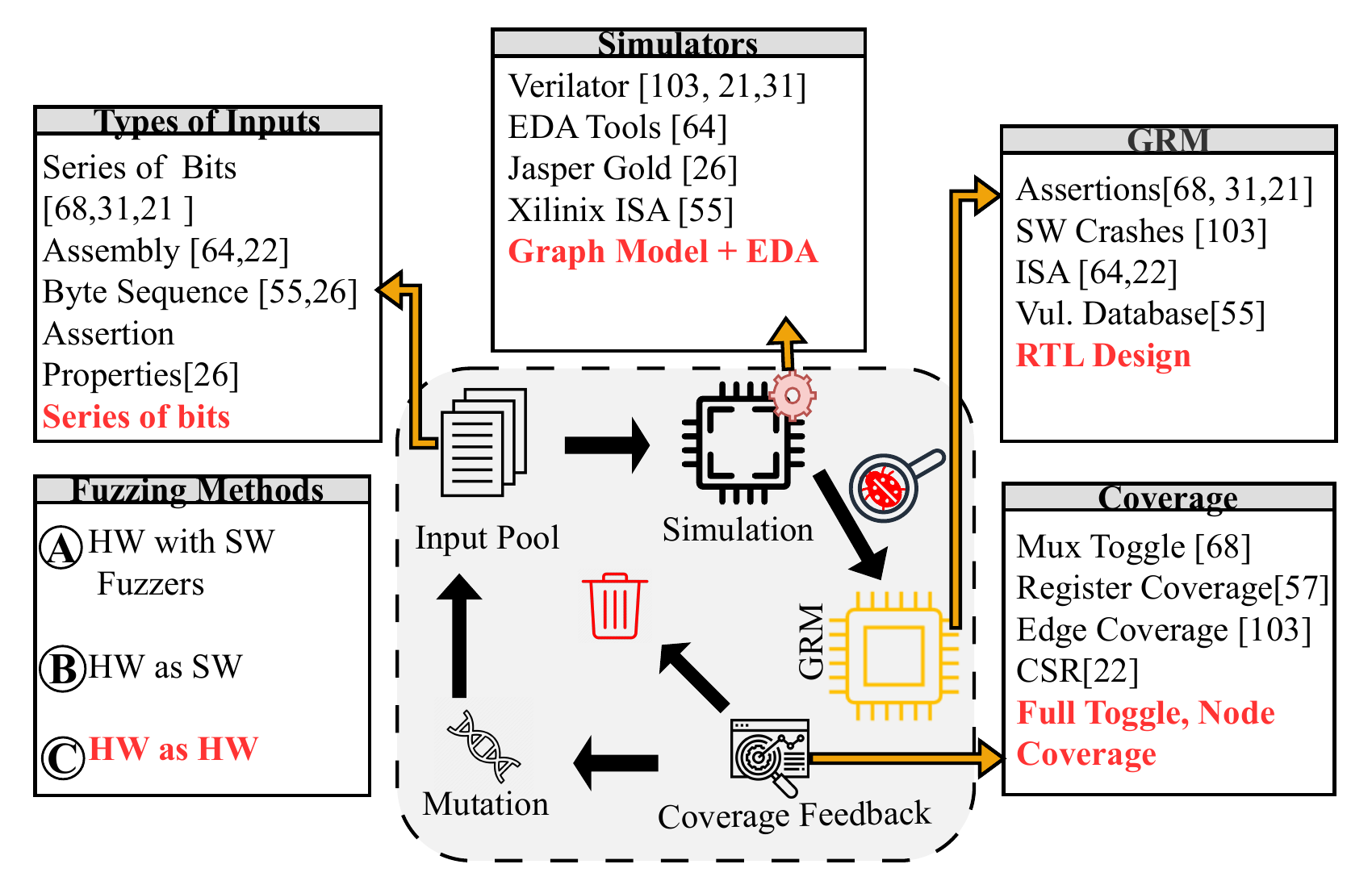} \vspace{-1em}
		\caption{Overview of hardware fuzzing, illustrating the selection of different parameters used in prior work and in our proposed GraphFuzz (highlighted in red)
		} \label{fig:hwfuzzfig}
		\vspace{-0.5em}
	\end{figure}

	Fuzzing techniques have proven to be efficient in detecting bugs and vulnerabilities both at the software and hardware levels of abstraction, as demonstrated by prior works \cite{ossfuzz,microsoftrisk}. To surpass the existing challenges in design verification, particularly in scalability for complex designs and automation, hardware fuzzing framework were proposed \cite{Wile'05,Clarke'12, Hicks'15, Guo17,Guo2017,Guo3017,Farahmandi,   bmc, Drzevitzky2010,Z3}. The hardware concept of fuzzing primarily involves: \textbf{\textit{1) random test case generation and mutations; 2) simulating the DUT (Device Under Test); and 3) analyzing for bugs or errors}} as illustrated in Figure \ref{fig:hwfuzzfig}. In hardware, the DUT represents the hardware at different levels of abstraction including architecture level, RTL level, and gate level. The core concept behind traditional fuzzing begins with the generation of \textbf{ random acceptable test case generations } (i.e., input stimuli) in tandem with the abstarction level chosen. 
	The DUT is simulated with these inputs to monitor its outcome, which is then verified against the expected output from the Golden Reference Model (GRM). However, when fuzzing large and complex designs, rather than relying solely on random inputs, a more efficient approach is to analyze the impact of these inputs on the DUT. Prior works have adopted code coverage as a feedback mechanism to the fuzzer, helping steer the fuzzer toward quickly exploring and covering the design space of the DUT as depicted in Figure \ref{fig:hwfuzzfig}. The existing works have adopted various simulators, coverage metrics, selection of Golden Reference Models (GRM), and input formats, as illustrated in Figure \ref{fig:hwfuzzfig}.

	Based on the chosen hardware level of abstraction and the fuzzing methodologies, the existing fuzzing frameworks can be classified into \textbf{ 1) Direction Adoption of Software Fuzzer for Hardware \textcircled{A} 2) Fuzzing Hardware as Software \textcircled{B} 3) Fuzzing Hardware as Hardware \textcircled{C}.} However, the limitation of \textbf{\textcircled{A}} \cite{Laeufer'18} is that hardware and software operate at different levels of abstraction, and unlike software, hardware does not inherently crash, making traditional software fuzzing techniques less effective \cite{Trippel'22}. In the case of \textbf{\textcircled{B}} \cite{Trippel'22}, treating hardware as software fails to account for intrinsic hardware characteristics, which are crucial for accurate bug detection \cite{Verilator}. Additionally, one of the main challenges with these approaches is extracting suitable coverage metrics—either from the translated hardware model or from the software fuzzer itself—since these metrics are not naturally aligned with the hardware's unique behaviors \cite{Trippel'22,Laeufer'18}. To circumvent these challenges, fuzzing hardware as hardware \textbf{\textcircled{C}} was proposed \cite{Kande'22,hybrid,borkar2024whisperfuzz}, which retains the inherent properties of the hardware during the fuzzing process. This approach has proven effective in detecting bugs and vulnerabilities while preserving the hardware's native form, making it more suitable for complex hardware designs. So far as mentioned earlier in \ref{intro}, the prior art proposed fuzzing at RTL \cite{Laeufer'18, Kande'22, Canakci'23,hybrid,borkar2024whisperfuzz}. While we outline the importance of fuzzing at gate-level in Section \ref{motivation}, we briefly  outline the fundamental characteristics that a hardware fuzzer should possess for effective bug detection.

	\begin{figure*}[h]
		\centering
		\includegraphics[width=1\linewidth]{
			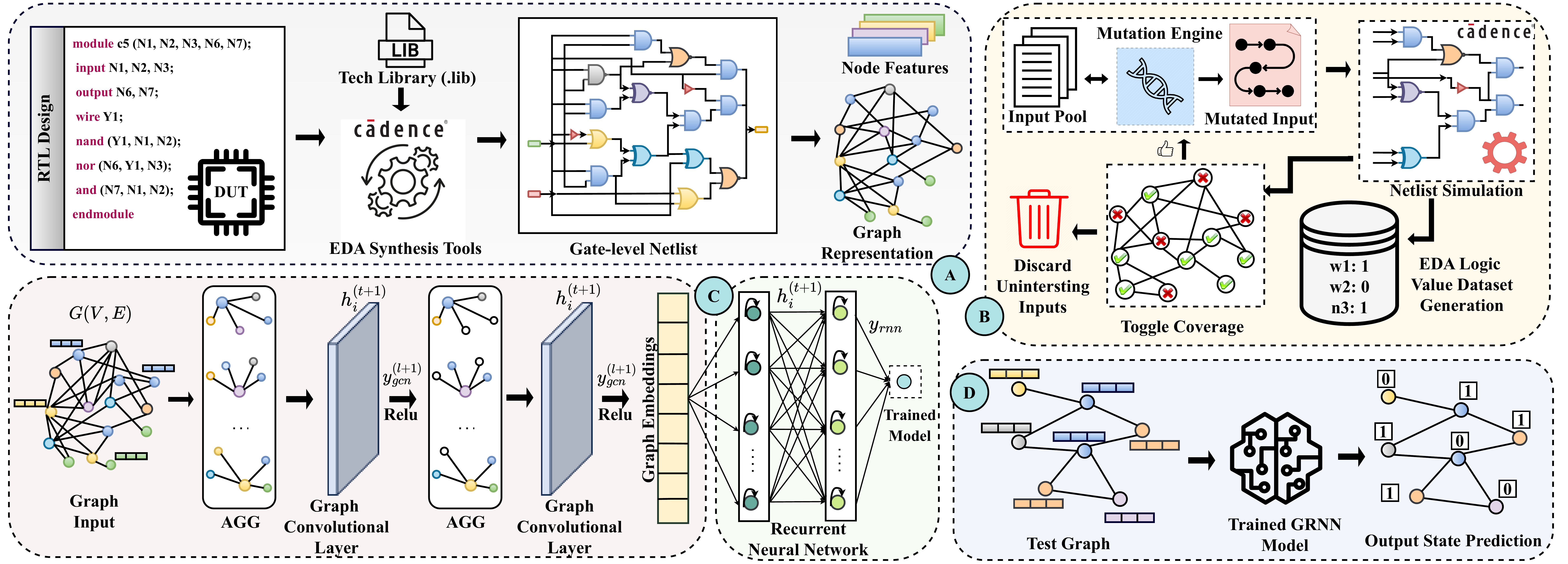
		} 
		\caption{Overview of our GraphFuzz Framework. \textcircled{A} Graph Generator (Section \ref{graphgen}), \textcircled{B} EDA Dataset Generation (Section \ref{edadataset}), \textcircled{C} GRNN Training (Section \ref{grnn}), \textcircled{D} GRNN Inference  
		} \label{fig:overview}
	\end{figure*}

	\paragraph{\textbf{Hardware Fuzzer Characteristics:}} The major challenge in designing a hardware fuzzer lies in determining the appropriate level of abstraction at which the hardware should be fuzzed. This could be at the architecture level, the RTL level, or the gate level, which is our main focus. Each level presents unique complexities and requires different strategies for effective fuzzing. Secondly, for a hardware fuzzer to be efficient, it must generate meaningful inputs that align with the chosen level of abstraction. Adopting inputs across different abstraction levels introduces difficulties due to fundamental differences in input formats and behaviors. Therefore, the input generation must be tailored to the specific abstraction level to ensure accurate and relevant testing. Moreover, the coverage metrics acting as feedback should be capable of capturing all the intrinsic characteristics of the selected abstraction level. These metrics are crucial for guiding the fuzzer, ensuring that critical parts of the design are sufficiently tested and that no corner cases are missed. Lastly, a suitable and trusted Golden Reference Model (GRM) must be available to validate the outputs during evaluation. 
	
	\subsection{Fuzzing Beyond RTL: The Need for Post-Synthesis Fuzzing} \label{motivation}
	Post-synthesis verification aims to ensure that the transition from RTL to gate-level design is flawless. While fuzzing can be applied at the RTL level, its effectiveness is limited because RTL simulations abstract away many of the details of the actual hardware. Synthesis tools \cite{Synopsys} perform numerous 
	optimizations to meet timing, area, and power constraints, such as collapsing gates, adding clock gating, or pruning redundant logic \cite{Nahiyan2019SecurityAwareFD}. While these optimizations improve the design's performance and efficiency, they can also introduce subtle functional changes that affect the correctness of the final circuit. For instance, during the synthesis process, EDA tools perform gate optimizations that introduce new data flows into the design, potentially altering its functionality and opening the door for vulnerability exploitation \cite{Dunbar2014DesigningTE}. One such bug was found in one of the recent fuzzing frameworks, Cascade \cite{cascade}, due to the logic synthesis error by the Yosys tool \cite{SymbiYosys}.
	
	Moreover, the mapping of RTL to the gate-level netlist is specific to the technology cell library being used. This library dependency can introduce variations in gate-level implementations. During synthesis, changes to the critical paths of a design may occur, leading to timing constraint violations. These timing issues, often involving setup and hold violations, arise when signals fail to arrive at registers in the correct sequence, causing incorrect data to be latched. Such issues are rarely caught during RTL simulations because RTL simulations do not model the gate-level timing accurately, especially those introduced by the specific library used during synthesis.In addition, for the concept of design for testability to facilitate the observability and controllability required for post-silicon validation of complex designs, scan chains are introduced. These test structures can also be exploited to attack the design by exposing internal data and control signals to an adversary, potentially leading to sensitive information leakage or tampering \cite{Sasan21,salmani2013}. Despite its importance, gate-level simulation is significantly slower than RTL simulation, as it involves modeling exact signal propagation, delays, and gate-level transitions. This makes exhaustive fuzzing at the gate level more challenging, as each test case requires more time to simulate. As we elaborate in Section \ref{ml_hs}, over the last decade, Machine Learning (ML) has emerged as a potential solution for addressing hardware security challenges. 
	ML techniques have been increasingly applied to various aspects of hardware design, verification, and security, offering advanced methods for detecting vulnerabilities, identifying anomalies, and improving the robustness of hardware systems.
	
	\vspace{-1 em}
	

	\section{ML for Hardware Verification}
	\label{ml_hs}

	It is estimated that verification engineers spend up to 70\% of their time on the verification phase in the modern IC design cycle \cite{vlsibook}. Machine Learning (ML) facilitates learning through training on data and evolutionary improvements, making it effective in handling both familiar and unseen data patterns. To reduce the manual intervention required from verification engineers, recent research trends focus on deploying ML techniques to automate various aspects of the verification process. By leveraging ML, tasks such as test case generation, bug detection, and coverage analysis can be automated, significantly reducing the time and effort required for verification while enhancing accuracy and coverage \cite{Selvakkumaran'03,Lee'08,Hamilton'17,Ying'18}. While applying Machine Learning (ML) to hardware verification, several key requirements must be met for effective implementation: \textbf{\textit{1) Definition of Problem Statement 2) Selection of Appropriate Hardware Abstraction Level 3) Selection of Processes to Automate 4) Generation of Unbiased, Realistic, and Comprehensive Data. }} In recent years there has been an exponential increase in the number of hardware verification works but not limited to the detection of trojans, securing designs against reverse engineering through ML, test case generation, side-channel analysis, and many more \cite{Cai'16,Ma'19,Haaswijk'18,Xie'18,Chan'16}. One such ML technique is the  Graph Learning algorithm, which is used in a diverse range of hardware security applications \cite{Selvakkumaran'03,Lee'08,Hamilton'17,Ying'18}.

	\vspace{-1 em}
	
	\subsection{Graph Learning for IC Security} \label{gleda}
	Graph Neural Networks (GNNs) have gained significant attention due to their superior performance in graph-based learning tasks, as highlighted in several studies \cite{Cai'18,Grover'16,Kahng'10,Yu'11,Cheng'95,Rozhin'21}. The success of GNNs in EDA is largely attributed to the fact that Boolean circuits can be naturally represented as graphs. A key factor driving the adoption of graph learning techniques \cite{Hamilton'17,Cao'15} over other methods is their ability to generalize, handle diverse topologies, scale efficiently, and be structurally aware, which serves as a \textit{primary motivation for us to choose graph learning in our work, GraphFuzz}. In contrast, traditional machine learning approaches, such as Deep Neural Networks (DNNs) and Convolutional Neural Networks (CNNs), have been widely applied across various EDA applications \cite{Cai'16,Chan'16,Chan'17,Jeong'10}, however, they require grid-based data or design representations  \cite{Hamilton'17,Cao'15}, limiting their use to specific topologies. Researchers have successfully applied GNNs to a range of electronic design automation (EDA) challenges, including detecting hardware Trojans (HTs) \cite{Yasaei2021GNN4TJGN,Yu2021HW2VECAG}, identifying IP piracy \cite{Baehr2019MachineLA}, reverse engineering gate-level netlists to name a few.

	
	\section{Hardware Fuzzing with Proposed GraphFuzz}
	\label{proposed}

	To address the challenges outlined in Section \ref{motivation}, we present a novel graph learning-based fuzzing framework, GraphFuzz, designed specifically for hardware fuzzing at the gate-level abstraction. \revision{The proposed work meets all the key requirements (outlined in Section \ref{ml_hs}) necessary for an ML-based hardware verification workflow}. 
	Our GraphFuzz unfolds in four steps as shown in Figure \ref{fig:overview}. Firstly, the DUT's gate-level netlist is modelled as an equivalent graph node model where each gate and interface ports (input, wire and ouput) is represented as a distinct node, followed by node encoding within the graph (Section \ref{graphgen} and \ref{nodeencode}) as illustrated in Figure \ref{fig:overview}\textcircled{A} and \ref{fig:nodeencode}. 
	This approach allows the retention of essential hardware computations directly at the gate level, preserving the structural and functional fidelity of the design. Secondly, we generate the EDA dataset required for the graph learning to learn the node features (Section \ref{edadataset}), as in Figure \ref{fig:overview}\textcircled{B}. Thirdly, these graph-based hardware model are trained with EDA datasets to generalize the intricate relationship between the nodes (Section \ref{grnn}) as in Figure \ref{fig:overview}\textcircled{C}. Lastly, the trained graph-based hardware model, having acquired knowledge of the DUT's characteristics which we term as NetGraph performs inferencing  as in Figure \ref{fig:overview}\textcircled{D},is fuzzed by the NetGraph Fuzzer (Section \ref{graphfuzzer}) to explore various graph regions to ensure the gate-level netlist adheres to high-level design intent captured in the RTL specifications. 
	GraphFuzz integrates smoothly with existing IC design and verification flows, 
	making it compatible with conventional hardware design methodologies while offering advanced, scalable, and accelerated gate-level hardware verification. 
	In the following section, we detail the steps involved in GraphFuzz fuzzing framework as outlined in Figure \ref{fig:overview}.
	

	\vspace{-1 em}
	
	\subsection{Graph Generator} \label{graphgen}

\begin{figure}[htb!]
		\vspace{-1em}
		\centering
		\includegraphics[width=0.7  \linewidth]{
			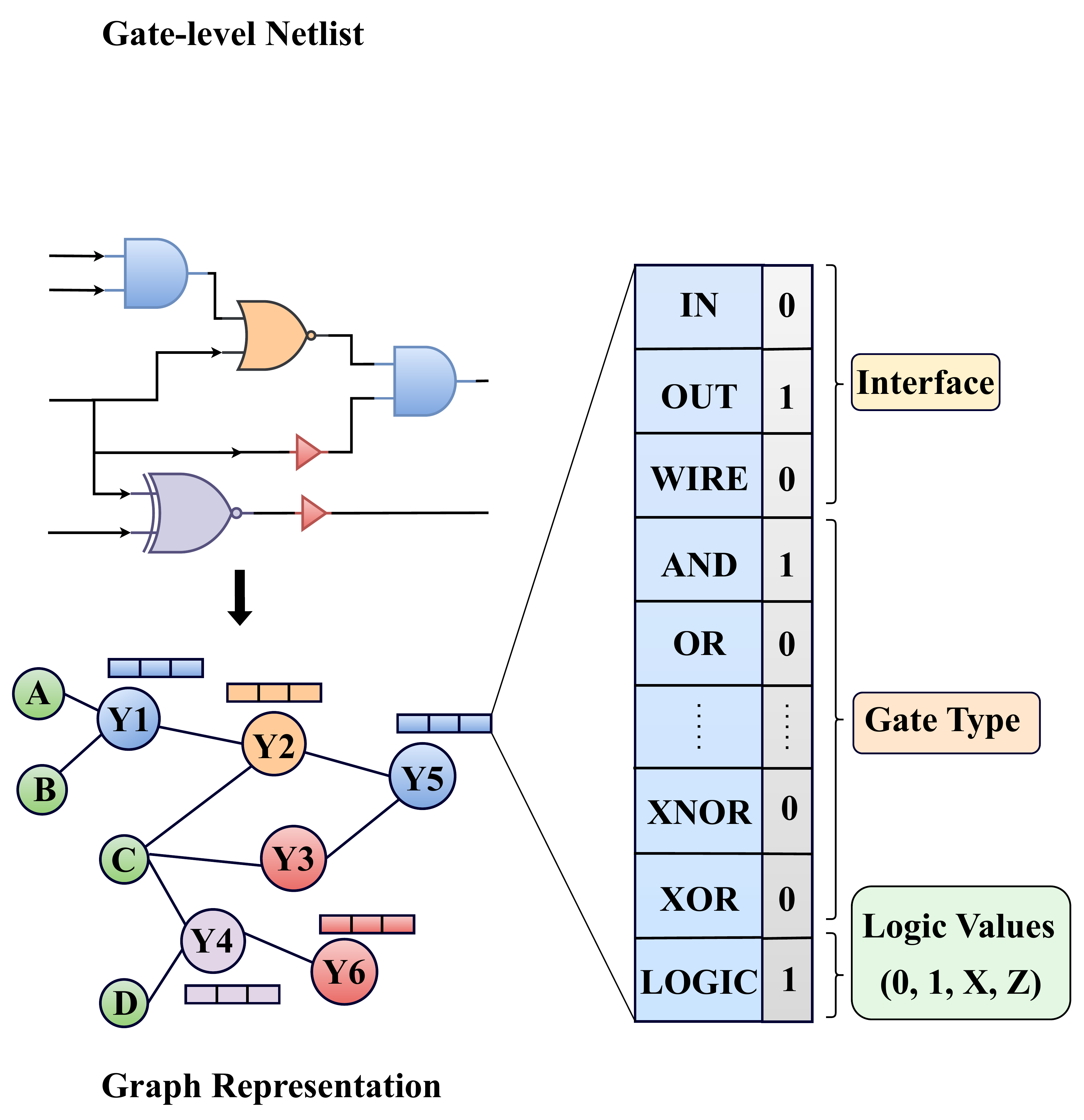} \vspace{-1em}
		\caption{Proposed Graph Node Encoding. GraphFuzz encodes three category of features : 1) Interface Type (Yellow) 2) Gate Type (Orange) and 3) Logic Value (Green)  
		} \label{fig:nodeencode}
		\vspace{-0.5em}
	\end{figure}

	\revision{To generate the graph representation of our DUT, we first extract the gate-level netlist from the RTL via the EDA synthesis tool.}
	The gate level netlist abstraction is modeled as graph nodes where the propagation of signals from one gate to another is akin to passing information along the edges of the graph as shown in Figure \ref{fig:overview}\textcircled{A} and Figure \ref{fig:nodeencode}. 
	At the gate-level simulation, the connectivity between gates and interface ports (inputs, outputs, and wires) is crucial for precise simulation and ensuring that the flow of logic values is correctly modeled and evaluated. 
	For effective graph-based representation, the feature vector of each node should encapsulate attributes analogous to those in EDA simualtion. To achieve this, a graph generator extracts the structural topologies (gates) and represents the interface ports (inputs, outputs, wires) of the netlist as graph nodes. While directionality between the ports and gates is essential for the simulation of digital circuits, the graph can be represented as undirected, provided that the logic values of each node are encoded as features that evolve over time, reflecting changes in the logic values of neighboring nodes. This approach allows the graph to capture the dynamic behavior of the circuit while simplifying its representation. Thus, we represent the circuits as a undirected graph $G = (V, E)$ consists of nodes (vertices) $V$ and edges $E$. The feature vector $f$ of each nodes are encoded in 2-D matrix, with a length of $k$ as shown in Figure \ref{fig:overview} and \ref{fig:nodeencode}. We outline the node encoding of our graph in Section \ref{nodeencode}.

	\vspace{-0.5 em}

	\subsection{Node Feature Encoding} \label{nodeencode}

	Our primary objective is to predict the logic values of nodes in the DUT, where each node's value changes dynamically based on its neighboring nodes translating to a graph-based learning problem. For this purpose, it demands accurate node embedding feature to distinguish between interface ports and gates. The feature vector for each node is designed to capture several key attributes.

	We present our node encoding feature and the corresponding graph 
	for a simple circuit in Figure \ref{fig:nodeencode}. Firstly, the feature vector encodes the interface type of each node, differentiating between various categories such as inputs, outputs, and internal connections (wire). This encoding is essential because the logic values of interface ports are crucial for understanding the flow of information between gates, as shown in Figure \ref{fig:nodeencode}. Secondly, the feature vector includes information regarding the gate type. This ensures that the functionality of the design is preserved in the graph representation, allowing the model to understand the role of each gate within the circuit. The third category of the feature vector encodes the logic value of the node at time $t$, reflecting the state of the node based on the logic values of its local neighbors similar to the different states of the circuit during simulation.  During the training phase, the gate type and interface type remain static. However, the logic value feature dynamically updates to reflect the varying input, wire, and output combinations, capturing changes over time. We extract a dataset required for the logic value feature vector and training as discussed in Section \ref{edadataset}. This allows the graph model to aggregate neighborhood values and simulate the circuit's evolving behavior, effectively mirroring the dynamic nature of circuit simulations. \revision{ For example, the node feature vector for the node Y5 in Figure \ref{fig:nodeencode} represents that the node is a output port (Interface) of a AND gate (Gate Type) and has the logic value (Logic Values) of 0 highlighted in Yellow, Orange and green respectively.} \revision{Given that now the node features are encoded the next step is to train the graph nodes with state-of-the-art graph learning algorithms as outlined in Section \ref{grnn} and Figure \ref{fig:overview} \textcircled{C}.}


	\vspace{-1 em}
	\subsection{Graph Recurrent Neural Network}
	\label{grnn}

	\revision{In this paper, the goal is to build a machine learning based fuzzer which detects vulnerabilities efficiently. And this task is similar to the time-series prediction as the input states keep changing overtime. With the ever-changing input states, the output also changes. Graph Recurrent Neural Networks (GRNNs) are effective in scenarios where graphs evolve over time. This facilitates a way to perform fuzzing with varying input and if the output is predicted as necessary or any vulnerabilities hinder the proper prediction. To achieve this we incorporate a graph convolutional neural network with recurrent units to add the memory element to them as represented in Figure \ref{fig:overview}\textcircled{C}. This helps the model learn through neighborhood aggregations as well as the previous states.} 
	
	
	Graph Neural Networks (GNNs) are a class of neural network models specifically designed to handle data structured as graphs. A graph $G = (V, E)$ is built from a netlist  as represented in Figure \ref{fig:overview}\textcircled{A}. Where nodes (vertices) $V$ are the gates (inputs, wires and outputs) and edges $E$ are the connections. Each node have features such as input data, type of gate and interface. The graph topology and the relationship between neighbors can be described using an adjacency matrix. It represents the connections between nodes.
	Each layer in the graph convolution network (GCN) is represented in the Figure \ref{fig:overview}\textcircled{C}.
	GNNs are updated by neighborhood aggregation (AGG) as represented in Figure \ref{fig:overview}\textcircled{C}, where the messages received from the neighboring nodes are computed and a learning transformation is applied to calculate the output. By updating, the GCN layers learn the features of each node in the graph. Through multiple layers of aggregation and transformation, GCNs learn complex relationships and patterns within the input graph and its node features.
	
	ReLU is used as a non-linear activation function between layers. The step by step process of training a GRNN model on the netlist graph representations to build the proposed GraphFuzzer. 
	The training algorithm follows the steps presented in Figure \ref{fig:overview}\textcircled{C}. The input to the training algorithm is the gate-level netlist which is represented in the Figure \ref{fig:overview}\textcircled{A} and processed in Figure \ref{fig:overview}\textcircled{B}. This gate-level netlist is given as an input to the $netlist\_to\_graph()$ function to convert the netlist into a graph (Line 3 to Line 7). The functionality of converting gate-level netlist to the graph $G(V,E)$ is represented in Figure \ref{fig:nodeencode}. Each gate (XOR, XNOR, AND, OR, etc,.) is converted into a graph node $V$. The features $Fea$ of the graph $G(V,E)$ is described with certain characteristics of the gate-level netlist such as interface (input, output or wire), gate type (XOR, XNOR, AND, OR, etc,.) and logic values (0,1,X, Z). Adjacency matrix $\hat{A}$ is defined by the connections between the gates in the circuit and it represents the connections between different nodes in the graph. The training process of the proposed graph recurrent neural network (GRNN) is defined from Line 8 to Line 12. The model training starts with passing the inputs to the graph convolutional layer (GCN). The inputs includes the generated graph dataset $G(V,E)$, the features of the graph $Fea$ and the normalized adjacency matrix $\hat A$. The definition of normalized adjacency matrix is presented in Equation \ref{eq2}.
	
	\begin{equation}
		\hat{A} = D^{-1/2} (A + I) D^{-1/2}
		\label{eq2}
	\end{equation}
	
	where $D$ is the degree matrix and $I$ is the identity matrix. \revision{Without normalization, the values of the node features might grow uncontrollably large as information is aggregated over multiple layers. This can lead to vanishing or exploding gradients during backpropagation, making training unstable. To avoid this we normalize the adjacency matrix $\hat{A}$.}
	
	The GCN layer operation with respect to the inputs as represented in Figure \ref{fig:overview}\textcircled{C} can be defined as follows:
	
	\begin{equation}
		y^{(l+1)}_{gcn} = ReLU(\hat{A}h^{(l)}W^{(l)})
		\label{eq1}
	\end{equation}
	
	Where $\hat{A}$ is the normalized adjacency matrix with added self-loops, $h^{(l)}$ is the node feature matrix at layer $l$, and $W^{(l)}$ is the weight matrix for layer $l$. 
	\revision{The weight matrix optimization is essential for training the GNN model. The weight matrix is initialized randomly or using a specific initialization technique, such as Xavier or He Normal initialization. During the forward pass, GCN layers aggregate information from neighboring nodes to update the node representations. Once the loss is computed, the GNN adjusts the weights through backpropagation and gradient descent, or its variants like Adam or RMSprop. Using the gradients, the model updates the weight matrix $W$ for each layer using an optimization algorithm. This forward pass and backpropagation process repeats over several epochs, with the GNN gradually adjusting its weights based on the training data and converging.}

	The ouput of the GCN layer is passed through the ReLU activation function and the output is defined in Equation \ref{eq1}. The output of the GCN is flattened and passed through the Long Short-Term Memory (recurrent unit) layer as represented in Line 11. 
	Recurrent Neural Networks (RNNs) are designed to handle sequential data by maintaining a hidden state that captures information from previous time steps as represented in Figure \ref{fig:overview}\textcircled{C}. The core idea is to use this hidden state to process sequences of inputs, allowing the network to learn temporal dependencies. At each time step 
	$t$, the RNN updates its hidden state $h_t$	and produces an output $y_{rnn}$ as represented in Figure \ref{fig:overview}\textcircled{C} based on the input $x_t$:
	
	\begin{equation}
		h_t = \Phi(W_h h_{t-1} + U_h x_t + b_h)
		\label{eq3}
	\end{equation}

	\begin{equation}
		y_{rnn} = W_y h_t + b_y
		\label{eq4}
	\end{equation}
	
	Where $W_h$, $W_y$, $U_h$ are the weights and $b_h$, $b_y$ are the biases. $\Phi$ is the activation function of the RNN model.
	
	
	The combined graph recurrent neural network defines each node $i$ in a graph at time $t$, the state update involves both graph convolution and recurrent components: 
	
	\begin{equation}
		h_i^{(t+1)} = \Phi(W_r h_i^{(t)} + U_r \sum_{j \in N(i)}h_j^{(t)} + b_r)
		\label{eq5}
	\end{equation}
	
	\begin{equation}
		h_i^{(t+1)} = RNN(h_i^{(t)}, GraphConv(h_i^{(t)}, N(i)))
		\label{eq6}
	\end{equation}
	
	Here, RNN represents the recurrent (LSTM) update mechanism and 
	GraphConv represents the graph convolution operation. $W_r$,  $U_r$ are the weights and $b_r$ is the bias. $h^{(i)}$ is the node feature matrix at layer $i$. $\Phi$ is the activation function of the RNN model. 
	
	\revision{The final output of the GRNN model is defined in Line 13. It refers to the output model trained on the input graph.}
	

	\subsection{EDA Dataset Generation} \label{edadataset}

	The effectiveness and accuracy of an ML model largely hinge on the quality and volume of the data it is trained on. However, acquiring a substantial amount of relevant, unbiased, and thorough data remains a challenge. In digital circuits, the logic value combinations of each node change dynamically over time as inputs vary, leading to an exponentially growing design space. As the size and complexity of the design under test (DUT) increase, the possible combinations of node values (inputs, wires, and outputs) multiply, making exhaustive simulation impractical. Considering this vast design space, it becomes crucial to focus on feeding the GNN model with data that represents realistic circuit behavior. To address this challenge, we leverage industry-standard EDA tools to simulate the hardware at the gate level using input seeds as in Figure \ref{fig:overview}\textcircled{B}. 
	
	These tools extract comprehensive node-level information across a wide range of input conditions, tracking how the logic values within the netlist evolve dynamically over time. This process enables the capture of realistic and time-varying logic states, providing insights into the functional behavior of the circuit at a granular level, which is encoded in feature vectors as described in \ref{nodeencode}. 
	To further enhance the model's efficiency in predicting node logic values, we implement an AFL-style mutation engine \cite{AFL'23} to generate a diverse set of input seeds. \revision{ Additionally, to train the GNN with diverse types of circuits and properties, we incorporate toggle coverage from the EDA tools as shown in Figure \ref{fig:overview}\textcircled{B}. This approach ensures that the circuit's behavior is comprehensively captured, enabling the model to learn and generalize more effectively for previously unseen input seeds.}

	\vspace{-1 em}
	\subsection{NetGraph Fuzzer} \label{graphfuzzer} 
	
	The trained netlist graph model, which encapsulates the behavioral characteristics and referred to as NetGraph (Figure \ref{fig:netgraphfuzzer} \textcircled{2}), is fuzzed to accelerate gate-level verification by utilizing the graph inference algorithm (Section \ref{infer}, in conjunction with EDA tools, as illustrated in Figure \ref{fig:netgraphfuzzer}. We now outline our NetGraph Fuzzer in the below section.

	\subsubsection{\textbf{Input Seed Generator:}}

	We now delve into the input seed generation process in the context of fuzzing the graph-based model (NetGraph) (Figure \ref{fig:netgraphfuzzer} \textcircled{1} ), aiming to explore various regions of the graph. In general, for the fuzzing process to be effective, it must generate input seeds in a format that aligns with the structure of the target entity. \revision{ Unlike prior works \cite{Kande'22,Canakci'23}, where the input seeds are at instruction level, since our DUT is represented at the gate-level abstraction, modeled as graph nodes, we concatenate all input pins and map the resulting bit vector in one test cycle as an input seed to the graph-based model} as shown in Figure \ref{fig:netgraphfuzzer} \textcircled{1}. This approach mirrors the way inputs are applied at the gate level of hardware abstraction, ensuring that the generated seeds are meaningful and relevant to the fuzzing entity. The input seeds are data files containing the bit vector as seeds, which is passed on to the input nodes of the graph for prediction of logic values for an accelerated fuzzing at the gate level.

	\subsubsection{\textbf{Test Suite Environment:}}
	
\begin{figure}[htb!]
		\centering
		\includegraphics[width=0.9  \linewidth]{
			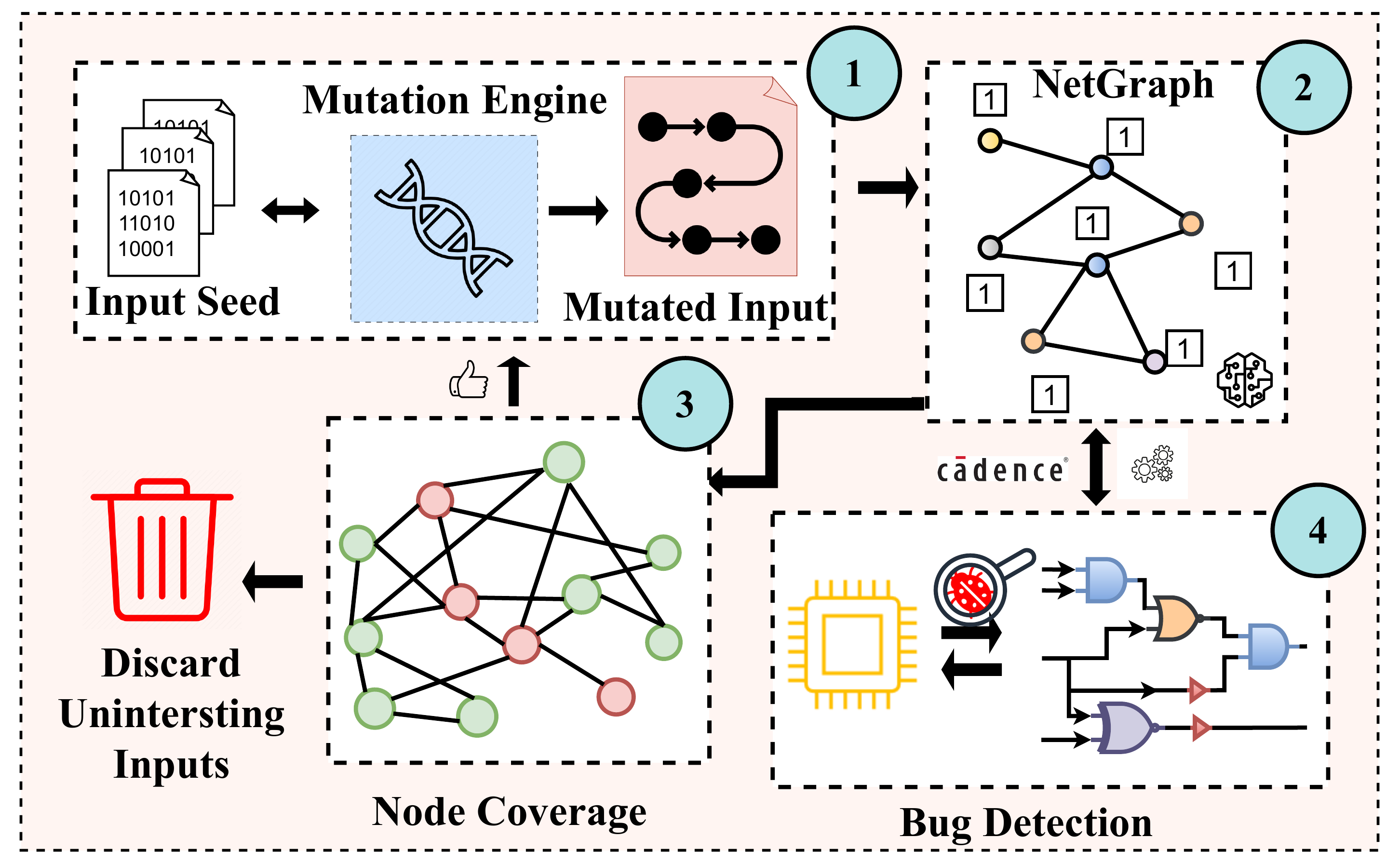} \vspace{-1em}
		\caption{Overview of Netgraph Fuzzer
		} \label{fig:netgraphfuzzer}
	\end{figure}

	The test suite generator is responsible for driving the graph-based model with the input seeds and mutating the input seeds based on the node coverage feedback. The initial set of seeds are generated through the seed generator
	and is mutated upon the node coverage feedback as shown in Figure \ref{fig:netgraphfuzzer} \textcircled{1} and \textcircled{4} respectively. The test suite environment mutates the interesting seeds based on the node coverage feedback, while discarding the uninteresting input seeds with less node coverage. This aids in steering the fuzzing process to trigger the various regions of graph nodes. The mutation operations are at the bit vector level to retain a meaningful input format to the graph nodes. For the mutation engine, we employ AFL-based mutation algorithm as traditionally used by majority of the hardware fuzzers.
	
	\subsubsection{\textbf{Coverage:}}
	\label{coverage_metrics}

	To ensure that the NetGraph Fuzzer thoroughly explores different regions of the graph, we utilize graph node coverage metrics, as illustrated in Figure \ref{fig:netgraphfuzzer} \textcircled{3}. The selected graph coverage metrics should correspond to those used in gate-level netlist verification, ensuring consistency between the graph model and the hardware-level abstraction. At the gate level, the primary coverage metric is toggle coverage, as described in Section \ref{edadataset}, which measures the toggling activity of gate-level elements during simulation.
	
	To maintain alignment between the chosen coverage metrics for graph nodes and the hardware-level gate-level netlist, \textit{ we apply four key graph metrics: Degree Centrality, Betweenness Centrality, Closeness Centrality, and Eigenvector Centrality}. These metrics are analogous to toggle coverage at the hardware level, ensuring that critical paths and nodes are comprehensively tested in both graph and gate-level abstractions. By leveraging these metrics, the NetGraph Fuzzer can effectively identify and explore potential vulnerabilities across all important regions of the circuit design.

	\textbf{Degree Centrality} is defined with the number of nodes connected to particular node, out of all the possible nodes in a given graph. We calculate the average degree centrality of a graph which is the mean of the degree centralities of all nodes in a graph. And it describes how well-connected each node is in the graph.
	
	\textbf{Betweenness Centrality} is defined with how often a node acts as bridge between two other nodes. The average betweenness in a graph represents represents the importance of node in flowing information between nodes. If the average is high, it indicates that many nodes have control over the information flow between the graph. If the average is low, it indicates that the graph is resilient to control points.
	
	\textbf{Closeness Centrality} measures how quickly a given node can access information from other nodes in the graph. Average closeness describes how close nodes are in a graph.
	
	\textbf{Eigenvector Centrality} measures the influence of a graph node not only in terms of number of connections but also the quality of connections it posses.

	\subsubsection{\textbf{GRNN Model Inference:}} \label{infer}

 The input to the inference algorithm is the test pool, mutated test samples generated using the mutation engine described in the Figure \ref{fig:overview}\textcircled{D} and the trained GRNN model. By passing the test mutated gate-level netlist through the $netlist\_to\_graph()$ function the test graph is generated $G_T(V, E)$. For each sample in the graph $v$ inference is performed to gather the predictions. The output predictions of GRNN model represented in Figure 
		\ref{fig:overview} \textcircled{D} can be defined as in Equation \ref{eq6}.

	\subsubsection{\textbf{Golden Reference Model:}} We use the RTL as the GRM for detecting discrepancies in the gate-level netlist as shown in Figure \ref{fig:netgraphfuzzer} \textcircled{4}. In addition we also use commercial EDA toolsto supplement the graph model for an effective bug detection as depicted in Figure \ref{fig:netgraphfuzzer}. By comparing the behavior and outputs of the gate-level netlist from against the RTL model, we can identify mismatches or deviations, ensuring that the synthesized gate-level implementation accurately reflects the intended design functionality captured in the RTL. This comparison helps in detecting bugs introduced during synthesis, optimization, or other transformations in the design flow, allowing for precise detection and debugging of any inconsistencies between the RTL and gate-level netlist.

	\section{Implementation}
	\label{implemtation}

	\begin{table*}[h!]
		
		\centering
		\caption{ {GRNN Architecture And Sampling Details} }
		\vspace{-1em}
		\label{tb1:grnn_architecture}
		\scalebox{0.8}{
			
			\begin{tabular}{|c|c|c|c|}
				\hline 
				
				\multicolumn{2}{|c}{\textbf{Architecture}} & \multicolumn{2}{|c|}{\textbf{Training and Sampling}}  \\ \hline
				\cline{1-4}
				
				\textbf{Total \# Layers} & 5 & \textbf{LSTM Output Layer}  & [\# output state predictions] \\
				\hline
				
				\textbf{GCN Depth} & 4 & \textbf{Activation} & ReLU  \\
				
				\hline
				\textbf{GCN Input Layer}  & [128, 256] & \textbf{Optimizer} & RMSprop\\
				\hline
				
				\textbf{GCN Hidden Layer} & [512, 256] & \textbf{Dropout} & 0.1   \\
				\hline
				
				\textbf{GCN Output Layer} & [256, 256] & \textbf{Learning Rate} & 0.001 \\
				
				\hline
				
				\textbf{Aggregation} & Attention aggregator & \textbf{Batch Size} & 128 \\
				& with multi-head concatenation &  &  \\
				
				\hline
				
				\textbf{LSTM Depth}  & 1 &\textbf{Weight Initialization } & HeNormal \\
				\hline
				
				\textbf{LSTM Input Layer} &  [256, 1, 256] & \textbf{Max \# Epochs} & 200 \\

				\hline
				
			\end{tabular}
		}
	\end{table*}
	
	The proposed GraphFuzz is compatible with the conventional industry standard IC design and verification flow enabling seamless verification of gate-level netlist. The majority of our components are designed using Python scripts. The toolchains used for the implementation of GraphFuzz is explained in this section.
	
	\paragraph{\textbf{Netlist and Graph Generation:}} We leverage the industry-standard EDA tool, Cadence Genus, to generate the gate-level netlist from a given RTL design. Cadence Genus supports a wide spectrum of HDL designs and integrates with various technology libraries to accurately map the RTL design to a gate-level representation. To facilitate the transition from a gate-level netlist to a graph suitable for machine learning models, we developed custom Python scripts. These scripts are responsible for extracting critical information such as gate types and interface ports (inputs and outputs) as discussed in \ref{nodeencode}, which are essential for building the graph. 
	
	\paragraph{\textbf{Dataset Generation:}} For the dataset generation required for graph training, we simulate the gate-level netlist generated by Cadence Genus using Cadence Xcelium. Cadence Xcelium is a standard EDA tool widely used in industry and academia for simulating hardware designs at various levels of abstraction, including RTL, gate-level netlists, and even transistor-level models. The seamless integration between the synthesis tool Genus and the simulation tool Xcelium enables a smooth flow from synthesis to simulation, ensuring that the hardware's behavior is accurately captured at the gate level. This process provides comprehensive data on the logic transitions of each gate and signal in the design, enabling us to effectively encode the logic value feature required for graph-based learning models. In addition, we developed custom TCL scripts for extracting the logic values of the nodes required for the graph model during the simulation. To extract the toggle coverage metric we leverage Cadence Integrated Metrics Center (IMC) and the feedback engine reads the coverage metrics through custom TCL and Python scripts. 
	
	\paragraph{\textbf{Seed Generator and Mutation Engine:}}  The seed generator is designed using custom python scripts to generate initial seeds required for the graph fuzzing. The test suite generation is designed seamlessly integrates with the seed generator and the feedback engine. The AFL mutation engine \cite{AFL'23} deployed using Python scripts mutates upon the feedback engine. 
	
	\section{Evaluation}
	\label{results}

	\begin{table*}[h!]
		
		\centering
		\caption{ {Graph Coverage Metrics of Different Hardware Designs} }
		\vspace{-1em}
		\label{tb2:coverage_metrics}
		\scalebox{0.9}{
			
			\begin{tabular}{|c|c|c|c|c|c|c|c|c|c|}
				\hline  
				\textbf{Dataset} & \textbf{\# of Inputs}  &  \textbf{\# of wires}   &   \textbf{\# of outputs}  &  \textbf{\# of Nodes}   &  \textbf{Degree Centrality} & \textbf{Betweenness} & \textbf{Closeness} & \textbf{Eigenvector} \\
				\hline
				AES &  257  & 3968  &  128 & 4353  &  8.08688 & 1.28409 &  0.00627 & 0.00393\\
				\hline
				C5315 & 178   &  690  &   123 &  991  & 0.00360  & 0.00455 &  0.13568 & 0.01098\\
				\hline
				DSP  &  34  &  2603  &  32  &  2669  & 0.00098  & 0.00158 &  0.14135 & 0.00174\\
				\hline
				C7552 & 207 & 927 & 108 &  1242& 0.00249  & 0.00375 &  0.14883 & 0.00888\\
				\hline
				S13207 & 65 & 1027 & 152 &1244  & 0.00265 & 0.00205 & 0.21042 & 0.01166\\
				\hline
				S5378 & 38 & 1029 & 49 & 1116  & 0.00333  & 0.00334 &  0.21161 & 0.02123\\
				\hline
				IBEX ALU & 202 & 917 & 166 & 1285  & 0.00267 & 0.00247 &  0.14951 &  0.00944\\
				\hline
				Rocket ALU & 133 & 1537 & 129 & 1799  & 0.00234 & 0.00219 & 0.20460 & 0.00603\\
				\hline
				or1200 ALU & 123   &  1905  &   38 & 2066    & 0.00188 & 0.00221 &  0.17421 & 0.00495\\
				\hline
				mor1kx ALU & 298 & 3098 & 116 & 3512 & 0.00094 &  0.00182 &  0.10697 & 0.00315\\
				
				\hline
				Ariane ALU  &  208  &  2217  &  65  & 2490   & 0.00168  & 0.00160 &  0.18221 & 0.00514\\

				\hline
			\end{tabular}
		}
	\end{table*}
	
	The experiments were conducted on a 48-core Intel Xeon processor with 512 GB of RAM, while the training of the graph-based models was performed using NVIDIA T4 GPUs, ensuring efficient processing and scalability across complex hardware designs.
	We evaluate GraphFuzzer on a diverse set of hardware designs to demonstrate its efficacy in learning complex node logic values, establishing a robust testbed for assessing its fuzzing capabilities. The selected designs encompass 1) ISCAS benchmark circuits, 2) IP peripherals \cite{opentitan}, and 3) Arithmetic Logic Unit (ALU) of open-source processors, including Ariane \cite{ariane}, mor1kx \cite{Morlkx}, IBEX \cite{RISC-V}, and Rocket \cite{RISC-V}, as referenced in Table \ref{tb1:grnn_architecture}. Unlike previous approaches primarily focused on processor designs, our fuzzer is capable of handling a broad spectrum of hardware architectures at a granular level, providing comprehensive coverage of both simple and intricate hardware characteristics. The designs selected for our proposed work include widely recognized processors and benchmark circuits, extensively used in hardware security research due to their flexibility and open-source nature. 
	
	The ISCAS benchmark circuits, known for decades in the hardware research community, are frequently employed for evaluating fault tolerance and security vulnerabilities in digital circuits. We also considered some of the IP peripherals such as AES \cite{opentitan} and DSP \cite{opentitan}. In terms of processors, IBEX is a 32-bit, low-power RISC-V core designed for embedded systems, making it a popular choice for research in secure, resource-constrained environments such as IoT \cite{RISC-V}. Rocket, a 64-bit in-order core from UC Berkeley, is highly configurable and commonly used in hardware security research for exploring microarchitectural modifications and security enhancements \cite{RISC-V}. Ariane, also know as \textit{cva6}, is a high-performance, 64-bit, out-of-order RISC-V core, suited for applications requiring both efficiency and computational power, making it ideal for high-performance security research\cite{ariane}. In addition, the popular OpenRISC based processor or1200 \cite{RISC} and mor1kx processor \cite{Morlkx} are open-source processors and are used in majority of the hardware verification works. These processors, along with the ISCAS circuits, form the backbone of security research, offering researchers a versatile platform to investigate vulnerabilities. 
	
	\subsection{Experimental Results}
	
\begin{figure}
		
		\includegraphics[width=0.5\textwidth, height= 5cm]{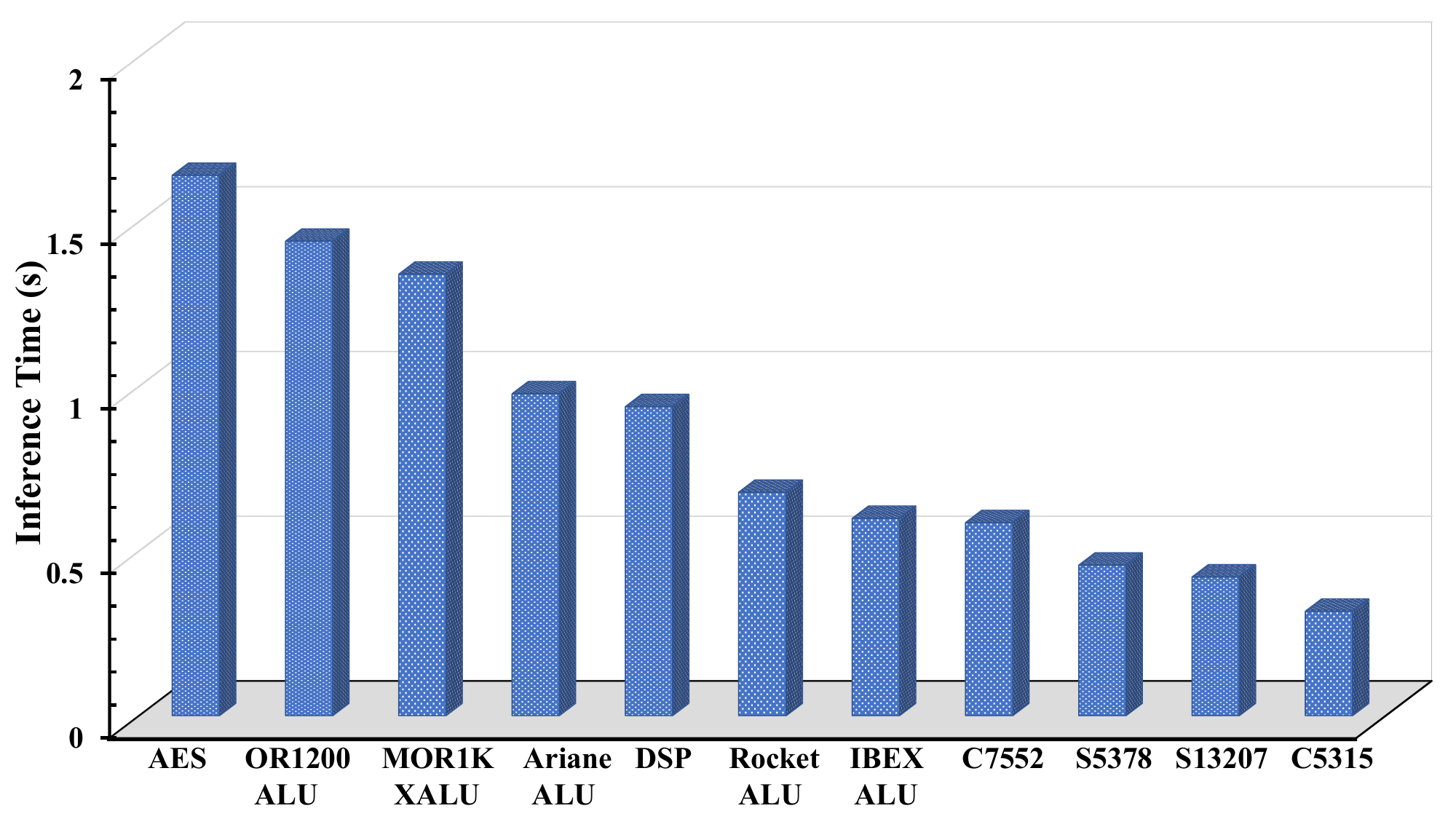}
		\vspace{-1em}
		
		\caption{Inference time of various architectures} 
		\label{fig:inference}
		
	\end{figure}

\begin{figure}
		
		\includegraphics[width=0.5\textwidth, height= 5cm]{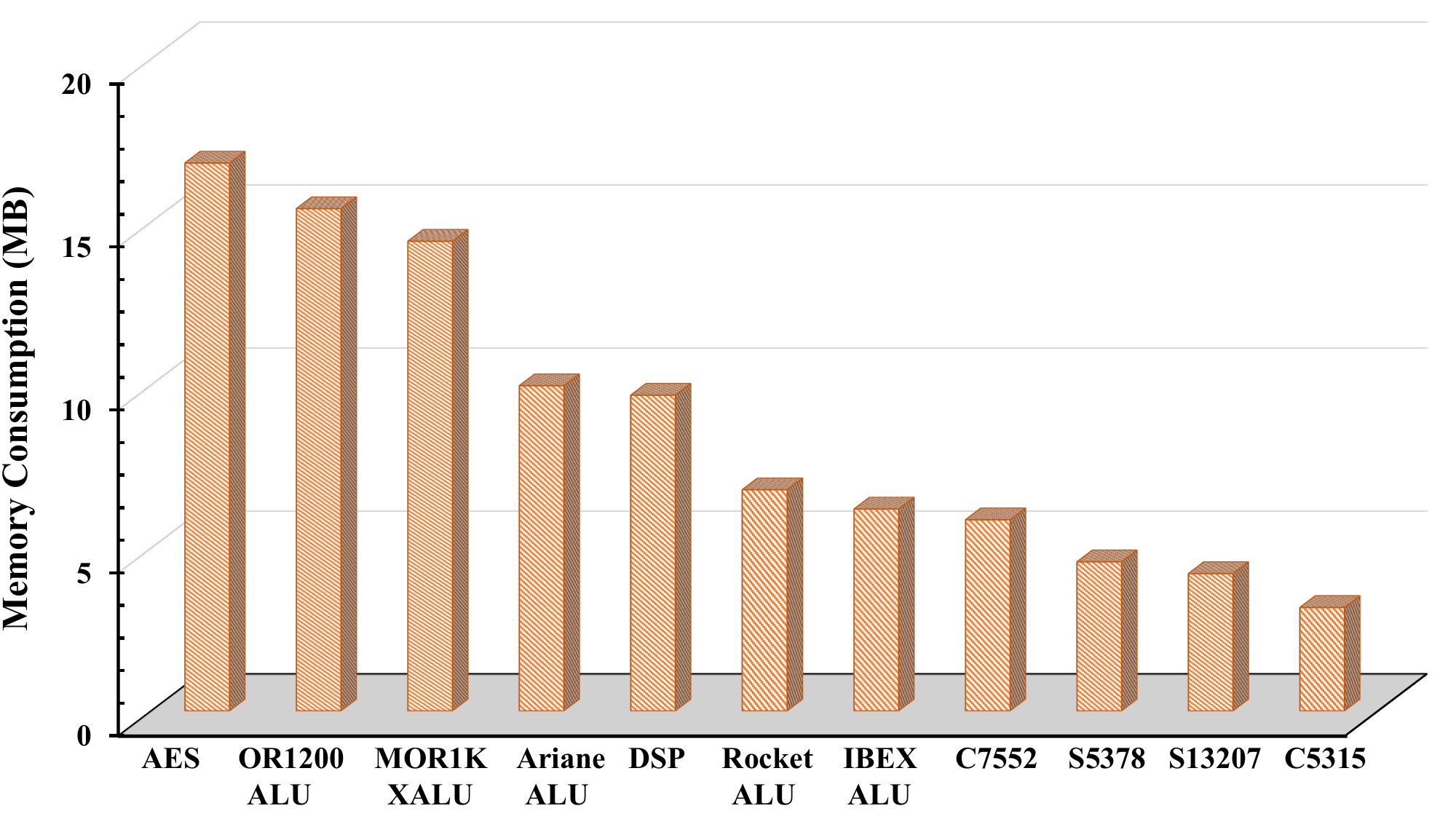}
		\vspace{-1em}
		
		\caption{Memory Consumption of various architectures} 
		\label{fig:memory}
		
	\end{figure}

	To build the GraphFuzzer we built a graph recurrent neural network (GRNN). In this section we provide the detailed description of the GRNN model as represented in Table \ref{tb1:grnn_architecture}. And evaluate the model's performance on various hardware designs as represented in Table \ref{tb3:accuracy}. We also present the inference timing and memory consumption of these hardware architectures as represented in Figure \ref{fig:inference} and Figure \ref{fig:memory}. The proposed GraphFuzzer has an architecture containing four graph convolutional layers (GCNs) and a recurrent unit. To avoid the vanishing gradient problem we employed Long Short-Term Memory (LSTM) layer. The LSTM incorporates the memory element and helps in making the time-series like predictions (prediction of output states to find vulnerabilities in the hardware circuit over changing inputs). The GCN input, hidden and output layer's dimensions are presented in Table \ref{tb1:grnn_architecture}. LSTM is given an depth of 1 by reshaping the output from GCN layer. The model is built with `ReLU' activation function, with `RMSprop' optimizer and the weights are initialized with `HeNormal' initialization. 
	
	The architecture of various hardware designs are presented in the Table \ref{tb2:coverage_metrics}. To describe the architecture of these designs we present the number of inputs, wires and outputs in each of the networks (which are converted as graph nodes to create a graph database). To further analyze these hardware designs we leverage certain graph coverage metrics such as degree centrality, betweenness, closeness and eigen vector. They are defined in the Subsection \ref{coverage_metrics}.

	AES, DSP, or1200 ALU, mor1kx ALU, and Ariane ALU are some of the largest hardware designs we considered, each with over 2,000 gates. C5315, IBEX ALU, and C7552 are among the smaller designs. AES has the highest degree centrality, indicating that the nodes in this graph are well-connected. On average, each node in the AES design is connected to more than eight other nodes (connections between inputs, wires, and outputs). mor1kx ALU and DSP have the lowest degree centrality, suggesting that the graph is poorly connected, which may affect the information transferred between nodes. The other designs have degree centralities ranging from 0.00234 to 0.00360, meaning that, on average, each node is connected to two or three others. AES also has the highest betweenness centrality, indicating that many nodes in the network play a key role in information transfer and serve as intermediates for shortest paths. In contrast, mor1kx ALU, Ariane ALU, and DSP have the lowest betweenness centrality, suggesting that these networks are decentralized. S5378 has the highest closeness centrality, meaning that its nodes are closer together, enabling faster information flow. AES, on the other hand, has the lowest closeness centrality, indicating that the nodes in its network are more distant, which increases communication time between them. S5378 also has the highest eigenvector centrality, suggesting that its nodes are well-connected to influential nodes. mor1kx ALU has the lowest eigenvector centrality, indicating that its nodes lack strong connections and have little influence over the data being transferred. Overall, AES and S5378 exhibit strong coverage metrics, showing that these networks have good connectivity between nodes, influence within the network, and support efficient data transfer.

	The performance metrics of different hardware designs such as accuracy, precision, recall and F1-score are presented in Table \ref{tb3:accuracy}. AES, S5378, and Ariane ALU have the highest accuracy of above 90\%, the precision, recall and F1-score also follow this pattern. Despite being one of the bigger hardware designs mor1kx ALU has the least accuracy and does not perform well. \revision{mor1kx ALU has low coverage metrics as discussed in Table \ref{tb2:coverage_metrics} which impacts the GRNN model performance. mor1kx ALU has lowest degree centrality, betweenness and eigenvector values which indicate that the network is poorly connected and nodes does not act as intermediate nodes. The nodes also face huge delays in information transfer between nodes. Which makes the nodes have limited access to information from the network. GRNN model learns from aggregating feature information from neighbouring nodes. But in the case of mor1kx ALU the model does not have enough scope to learn from neighbours and has less information to aggregate during training. This results in poor feature representation and lower accuracy.}
	Measures such as hyperparameter tuning is performed to improve the performance of this design but the improvement is minor. DSP, C7552 and Rocket ALU are some of the other designs which have an accuracy about 70\%. \revision{As they also have lower coverage metrics, the GRNN training is effected.}
	The performance metrics show that the prediction capabilities of the GRNN models trained on various hardware designs and their capability to predict the vulnerabilities of circuit without performing the manual hardware fuzzing task. Most of the hardware designs perform well on the inference task. Especially designs such as AES and S5378 which possess good connectivity between nodes as represented by the coverage metrics in Table \ref{tb2:coverage_metrics}. This improves the GRNN model's training process on these specific designs, leading to higher performance.

	\begin{table}[h!]
		
		\centering
		\caption{ {Performance Metrics of the Proposed GraphFuzz Model With Respect to Different Architectures} }
		\vspace{-1em}
		\label{tb3:accuracy}
		\scalebox{0.85}{
			
			\begin{tabular}{|c|c|c|c|c|}
				\hline  
				\textbf{Dataset} & \textbf{GRNN} &  \textbf{Precision} & \textbf{Recall} & \textbf{F1-score} \\
				
				&\textbf{Accuracy  (\%)}& \textbf{(\%)} & \textbf{(\%)} & \textbf{(\%)} \\
				\hline
				AES &  92.52 & 92.49 &  92.53 & 92.42\\
				\hline
				C5315 & 79.82  & 79.86 &  79.84 & 79.82\\
				\hline
				DSP  &  70.24  & 70.24 &  70.25 & 70.25\\
				\hline
				C7552 & 72.77  & 72.76 &  72.77 & 72.78\\
				\hline
				S13207 & 76.81 & 76.84 &  76.85 & 76.85\\
				\hline
				S5378 &  96.04  & 96.02 &  96.04 & 96.04\\
				\hline
				IBEX ALU  & 79.75 & 79.77 &  79.76 & 79.69\\
				\hline
				Rocket ALU &  73.82 & 73.86 &  73.86 & 73.87\\
				\hline
				or1200 ALU & 83.14  & 83.16 &  83.15 & 83.14\\
				\hline
				mor1kx ALU &  69.34 & 69.35 &  69.34 & 69.37\\
				
				\hline
				Ariane ALU  &   95.38  & 95.39 &  95.36 & 95.38\\

				\hline
			\end{tabular}
		}
	\end{table}

	The inference latency for various hardware design models is shown in Figure \ref{fig:inference}. The designs are arranged from highest to lowest latency. Larger circuits, such as AES, or1200 ALU, and mor1kx ALU, exhibit higher latencies. Specifically, AES has an inference latency of 1.64 seconds, or1200 ALU has a latency of 1.44 seconds, and mor1kxALU has a latency of 1.34 seconds. In contrast, smaller circuits such as S5378, S13207, and C5315 have lower latencies. S5378 has an inference latency of 0.457 seconds, S13207 has a latency of 0.421 seconds, and C5315 has a latency of 0.317 seconds.
	
	
	
	The memory consumption of various trained hardware design models is illustrated in Figure \ref{fig:memory}. The designs are arranged from highest to lowest memory usage. The AES, or1200 ALU, and mor1kx ALU circuits, being larger, consume more memory. Specifically, AES consumes 17 MB, or1200 ALU consumes 15 MB, and mor1kx ALU consumes 12 MB. On the other hand, the S5378, S13207, and C5315 circuits exhibit lower memory usage, with S5378 consuming 5 MB, S13207 consuming 4.5 MB, and C5315 consuming 3 MB.
	
	\begin{table*}[h!]
		
		\centering
		\caption{Comparison with existing HW Fuzzing Frameworks  }
		\vspace{-1em}
		\label{table:prior}
		\scalebox{0.75}{
			\begin{tabular}{|l|l|l|l|l|l|l|}
				\hline
				\textbf{Framework } & \textbf{Fuzzer Type} & \textbf{Input} & \textbf{Simulator} & \textbf{Coverage Metric} & \textbf{Target Design } & \textbf{Evaluation}    \\
				\hline
				
				RFUZZ \cite{Laeufer'18}  & HW Fuzzer (1) & Series of bits & Any & Mux Toggle & Peripherals, & Assertion  \\
				&&&& &RISC-V &\\
				
				\hline
				DifuzzRTL \cite{Hur'21}  & HW Fuzzer  & Assembly & Any & Register Coverage & RISC-V CPU & GRM  \\
				\hline
				Trippel et al \cite{Trippel'22} & SW AFL Fuzzer  & Byte Sequence & Verilator & Edge Coverage & AES,HMAC  & SW crashes  \\
				&&&& &KMAC, Timer & \\
				\hline
				HyperFuzzer \cite{hyperprop}  & SW AFL Fuzzer  & Series of bits & Verilator & High-level & SoC  & Assertion \\
				\hline
				DirectFuzz \cite{direct} & HW Fuzzer  & Series of bits & Verilator & MUX & Peripherals, & Assertion  \\
				&&&& &RISC-V& \\
				\hline
				TheHuzz \cite{Kande'22}  & HW Fuzzer  & Assembly  & Synopsys VCS & FSM, Branch,toggle, & RISC-V  & GRM  \\
				& & & & conditional& &  \\ 
				\hline
				Processor Fuzz \cite{Canakci'23} & HW Fuzzer  & Assembly & Verilator & Control path& RISC-V  &GRM \\
				&&& &  register, ISA-tranistion& &   \\
				\hline
				SoCFuzzer \cite{Hossain'23}  & HW Fuzzer  & Byte Sequence & Xilinx ISA & Randomness, target & SoC & Database  \\
				&&&& output, input coverage& &  \\
				\hline
				HyPFuzz \cite{hybrid} & HW Fuzzer   & Assertion Cover properties,   & Jasper Gold, & FSM, Branch,toggle, & RISC-V  & GRM     \\
				&& Byte Sequence & Synopsys VCS&conditional & &  \\
				\hline
				
				\textbf{Our proposed, GraphFuzz} & \textbf{HW Fuzzer} & \textbf{Series of bits} & \textbf{Graph Model,Cadence Xcelium} & \textbf{Node Coverage, Toggle} & \textbf{Netlist of ISCAS,} & \textbf{RTL}  \\
				&&&& & \textbf{AES, DSP, }& \\
				&&&& & \textbf{ALU of RISC-V \& }& \\
				&&&& & \textbf{OpenRISC}& \\
				
				\hline

				
			\end{tabular}
		}
	\end{table*}

	\subsection{Bugs Reported}
	In this section, we report the bugs found by our GraphFuzz, at the gate-level netlist of our selected designs. We first report the bugs for the ISCAS benchmark circuits, followed by the IP blocks in the below sections:
	
	\subsubsection{\textbf{C5315, C7552, S13207, S5378:}} Through comprehensive fuzzing experiments utilizing the NetGraph fuzzer on the selected designs in our study, we observed no discrepancies or bugs between the outputs derived from the RTL simulations and those generated by the NetGraph model. This result suggests a high degree of functional alignment between the RTL-level description and the gate-level abstractions generated by the EDA tools.

	\subsubsection{\textbf{mor1kx ALU:}}The bug orchestrated in the incorrect assignment of carry flag node for some instruction as detected by \cite{Kande'22}. Our GraphFuzz model achieved an approximate accuracy of 70\% in predicting the node value associated with the carry flag, as this bug manifests at the RTL level of abstraction. To evaluate the model's performance, we introduced mutations to the bug-triggering input  (i.e bit vectors for the sub instruction) and calculated the average accuracy across these variations. This bug leads to incorrect logic operations failing to meet the design specifications.
	
	\subsubsection{\textbf{or1200 ALU:}}Our GraphFuzz model was also successful in detecting the incorrect assignment of the overflow flag, achieving an average accuracy of 82.7\% across a range of test case scenarios, consistent with findings reported in previous works \cite{Kande'22}.
	
	\subsubsection{\textbf{Gate Delay Bugs:}} During the synthesis process, EDA tools introduce gate delays within the netlist, leading to transient inaccuracies in output calculations prior to the stabilization of the correct result. In our comparative analysis of graph learning predictions and synthesized netlist outputs against RTL simulations, we observed significant discrepancies in AES, DSA and in the ALU's of or1200, mor1kx, Rocket, IBEX. The bugs stems as multiple clock cycles exhibited erroneous or mismatched outputs due to propagation delays in the gate-level design, which were absent in the idealized RTL simulation. These delays are a consequence of real-world signal propagation and gate-level timing, which affect the accuracy of the output during the initial clock cycles until the system reaches a steady state. This highlights the critical impact of gate-level timing on design verification and the limitations of RTL simulation in capturing these phenomena.

	\section{Related Works}
	
	The prior works attempted to fuzz the hardware as outlined in Table \ref{table:prior} at different levels of abstraction. RFuzz \cite{Laeufer'18} was the work proposed in hardware aiming to  fuzz the RTL design by direction adoption of SW Fuzzers and uses mux toggle as the coverage metrics. However, it fails to scale and capture bugs. In contrast to RFuzz, Trippel \textit{et al.} \cite{Trippel'22} proposed fuzzing hardware-like software rather than porting software fuzzers directly on the hardware designs. While this approach is promising, these translated HW models do not support HDL constructs and account for innate hardware characteristics. In addition, the coverage metrics extracted do not apply for harwdare abstraction. Conversely, works \cite{hyperprop} proposed Hyperproperties,  which are higher-level properties that describe security policies by comparing the behavior of instances of a system. Unfortunately, these require human intervention and design knowledge and often fails for large Complex SoC designs. Later, few works \cite{Kande'22,hybrid,Canakci'23} proposed fuzzing hardware in its innate hardware level of abstraction as outlined in Table \ref{table:prior}. These fuzzing frameworks are highly efficient in detecting hardware bugs, as they are designed to account for the inherent characteristics of hardware systems. The coverage metrics selected for these frameworks are tailored to align with the traditional IC design flow. However, none of the existing frameworks fuzz at gate-level (i.e netlist) to find bugs.
	
	In contrast, our proposed GraphFuzz 1) supports conventional hardware design and verification methods, 2) captures intrinsic hardware characteristics using comprehensive EDA data, 3) accelerates hardware verification at the gate-level, 4) effectively detects bugs at the gate-level, and 5) does not require extensive design knowledge or expertise in hardware design.

	\section{Discussion and Limitations}
	\label{discussion}

	\paragraph{\textbf{RTL and Netlist Accessibility:}} GraphFuzz relies on access to the RTL or gate-level netlist, similar to prior fuzzing techniques. In modern IC design and verification flows, verification engineers typically have access to one or both of these representations. However, adversaries can exploit reverse engineering techniques to extract the gate-level netlist from a physical chip. Our GraphFuzz can seamlessly integrate with GNN-RE, a graph learning-based reverse engineering framework for extracting gate-level netlists, enhancing its applicability in scenarios where direct access to design files is unavailable. This synergy enables comprehensive fuzzing even when the original design information is obscured or unavailable. 
	
	\paragraph{\textbf{FPGA Designs:}} The existing hardware fuzzing frameworks typically depend on EDA simulation tools such as Synopsys VCS and Cadence Xcelium, making them unsuitable for fuzzing FPGA-emulated designs. For example, tools like DifuzzRTL \cite{Hur'21} and RFuzz \cite{Laeufer'18} primarily focus on fuzzing processor designs through FPGA simulations. In contrast, GraphFuzz can be extended to fuzz FPGA-based designs by extracting the gate-level netlist directly from FPGA toolchains and modeling the graph nodes with relevant features. Moreover, logic values can be seamlessly extracted from the FPGA-emulated designs, enabling the same graph learning and fuzzing techniques used for traditional netlist designs to be applied to FPGA-specific environments. This extension ensures that GraphFuzz remains versatile and adaptable across various hardware platforms, including FPGA-based systems.

	\paragraph{\textbf{Fuzzing Temporal Characteristics:}}To this end, GraphFuzz's graph model is primarily designed to encode the functional properties of the hardware design, but it does not currently account for temporal characteristics, such as timing violations. However, GraphFuzz can be extended to incorporate temporal characteristics within the graph model, enabling it to detect timing vulnerabilities.
	
	\paragraph{\textbf{EDA Optimizations:}}
	In our analysis of hardware vulnerabilities, we observed that optimizations introduced during synthesis, such as merging sequential instances, ungrouping hierarchical modules, and bitwidth mismatches, can create significant security risks that are difficult to detect using conventional verification methods. For instance, merging sequential elements, such as flip-flops in DSP and cryptographic units, facilitates adversaries' ability to exploit shared states and timing variations, leading to side-channel attacks or system-wide failures. Similarly, ungrouping hierarchical IP cores in SoC designs can expose internal logic, leading to privilege escalation and unauthorized access to secure components. Bitwidth mismatches, as identified in the or1200 processor ALU, result in incorrect calculations and lost bits, further compromising design integrity.
	
	Additionally, buffer optimizations—where buffers are inserted or removed to improve performance—can introduce timing vulnerabilities that open up attack vectors for inducing faults or extracting sensitive data via side-channel attacks. These vulnerabilities highlight the need for more advanced verification techniques. The types of bugs discussed here cannot be fully detected by our current graph-based hardware fuzzing approach GraphFuzz alone, as they require integration with EDA synthesis tools to account for gate-level design rules and signal integrity issues introduced during the synthesis process. By augmenting graph learning with synthesis-aware analysis, these complex bugs could be identified more effectively.

	

	\section{Conclusion}
	\label{conclusion}
	
	Hardware vulnerabilities have been steadily increasing over the years, and current hardware verification methods are struggling to keep up, particularly in the context of large-scale, complex designs. The scalability challenges faced by traditional verification techniques underscore the urgent need for more advanced approaches. In response to this, we present GraphFuzz, the first graph-based hardware fuzzing framework specifically designed for gate-level verification. GraphFuzz introduces a novel approach to hardware fuzzing by leveraging graph learning techniques to accelerate the verification process at the gate level. Our framework was able to detect bugs in real-world IP blocks. Our GraphFuzz can ease the gate-level simulations without much human interventions and efforts leveraging state-of-the-art hardware verification technique.

	\bibliographystyle{IEEEtran}

	\bibliography{ref,reference,references}

\vspace{12pt}

\end{document}